\documentclass[5p,times]{elsarticle}

\def\baas{\ref@jnl{BAAS}}               

\usepackage[colorlinks]{hyperref}
\usepackage{amsmath}
\usepackage{amssymb}
\usepackage{amsfonts}
\usepackage{fontawesome}
\usepackage{float}
\usepackage{xcolor}
\newcommand{\tens}[1]{\mathsf{#1}}

\journal{Journal of \LaTeX\ Templates}

\usepackage{xspace}
\newcommand{\scarlet}{{\tt scarlet}\xspace}
\newcommand{\galsim}{{\tt galsim}\xspace}
\newcommand{\Y}{{\bf Y}}
\newcommand{\X}{{\bf S}}
\newcommand{\tH}{{\bf H}}
\newcommand{\A}{{\bf A}}
\newcommand{\bY}{{\bf \bar{Y}}}
\newcommand{\bX}{{\bf \bar{S}}}
\newcommand{\bH}{{\bf \bar{H}}}
\newcommand{\bA}{{\bf \bar{A}}}
\newcommand{\bN}{{\bf \bar{N}}}
\newcommand{\Xp}{{X_{l_1}}}
\newcommand{\Yq}{{Y_{l_2}}}
\newcommand{\xii}{{x_{m_1}}}
\newcommand{\yj}{{y_{m_2}}}
\newcommand{\xk}{{x_{k_1}}}
\newcommand{\yl}{{y_{k_2}}}
\newcommand{\rect}{{\mathrm{rect}}}
\newcommand{\sinc}{{\mathrm{sinc}}}
\newcommand{\red}[1]{{\color{red}{#1}}}
\newcommand{\hide}[1]{}

\biboptions{authoryear}

\begin{document}

\begin{frontmatter}

\title{Joint survey processing: combined resampling and convolution for galaxy modelling and deblending}

\author[1]{R\'emy Joseph}
\author[1,2]{Peter Melchior}
\author[3]{Fred Moolekamp}
\address[1]{Department of Astrophysical Sciences, Princeton University, Princeton, NJ 08544, USA}
\address[2]{Center for Statistics \& Machine Learning, Princeton University, Princeton, NJ 08544, USA}
\address[3]{School of Physics and Astronomy, Rochester Institute of Technology, Rochester, NY 14623, USA}

\begin{abstract}
     We present an extension of the multi-band galaxy fitting method \scarlet which allows the joint modeling of astronomical images from different instruments, by performing simultaneous resampling and convolution.  We introduce a fast and formally accurate linear projection operation that maps a pixelated model at a given resolution onto an observation frame with a different point spread function and pixel scale. We test our implementation against the well-tested resampling and convolution method in \galsim on simulated images mimicking observations with the {\it Euclid} space telescope and the {\it Vera C. Rubin} Observatory, and find that it reduces interpolation errors by an order of magnitude or morecompared to \galsim default settings. Tests with a wide range of levels of blending show more accurate galaxy models from joint modeling of {\it Euclid} and {\it Rubin} images compared to separate modeling of each survey  by up to an order of magnitude. Our results demonstrate, for the first time, the feasibility and utility of performing non-parametric pixel-level data fusion of overlapping imaging surveys. All results can be reproduced with the specific versions of the codes and notebooks used in this study.
\end{abstract}

\end{frontmatter}
\section{Introduction}
\label{Sect:intro}

    Every large-scale space or ground-based survey leads to a wide variety of data products, each one with characteristics unique to the instrument and survey. For instance, the {\it Vera C. Rubin} Observatory Legacy Survey of Space and Time (LSST) \citep{Ivezic2019} will provide multiband observations down to r-band magnitude 27.5 over 18,000 deg$^2$. The {\it Nancy Grace Roman} Space Telescope \citep{Spergel2015, Akeson2019} will produce high resolution images in seven bands in the visible and near-infrared range over 2,000 deg$^2$ . The {\it Euclid} mission \citep{Laureijs2011} will produce wide-band high resolution imaging over 15,000 deg$^2$ also in the optical and near-infrared range. While each of these surveys are expected to yield unprecedented scientific results for the astronomical and cosmology community, we should also recognize that they are highly complementary. What the {\it Euclid} survey lacks in sensitivity and spectral resolution in the visible range, LSST can compensate with deep multi-band observations. Conversely, what LSST's images, taken with the {\it Simonyi} Survey Telescope, lack in spatial resolution, the two space telescopes {\it Roman} and {\it Euclid} can provide images with $\sim 2$ times higher pixel resolution.
    
    Several studies  have addressed the complementary nature of these surveys \citep{Capak2019, Eifler2019, Rhodes2017, Rhodes2019, Scolnic2018, Chary2019-kd, Chary2020-rr}, including estimated gains from combining observations at the catalog or at the pixel-level, for specific science cases, such as galaxy shape measurement \citep{Schuhmann2019}, weak lensing and galaxy clustering \citep{Eifler2020}, photometric redshift estimates \citep{Graham2020}, observations of tidal stellar streams \citep{Laine2018}, exoplanet searches \citep{Bachelet2019, Street2018} and solar system object tracking \citep{Snodgrass2018}.
    
    To jointly model observations at different resolutions at the pixel level, we could choose to employ parameterized models, which potentially renders changes of resolution or convolution with the point spread funstion (PSF) to be analytically expressible. But parameterized models do not fully reflect the variety of galaxy morphologies. While that can be acceptable for ground-based surveys, it becomes a noticeable shortcoming for observations from space, where intricate structures are visible in most galaxies.
    For non-parameteric methods, we need to build a resampling model that accounts for the difference in pixel size and positions between observations, the difference between the telescopes PSFs and the instruments' pixel integrations. The problem of resampling is well studied in the astronomical literature, and a number of methods exist, e.g. for forced photometry \citep{Merlin2016-oz}, for upsampling of undersampled images \citep{Bertin2011, Ngole2015, Rowe2011}, and general-purpose image resampling \citep[][BG hereafter]{Bernstein2014}.
    
    In this paper we implement a resampling scheme with minimal assumptions and approximations. We then integrate it into the galaxy modeling framework \href{https://github.com/pmelchior/scarlet}{\faGithub \scarlet}\footnote{https://github.com/pmelchior/scarlet} \citep{Melchior2018}, which specialises in multi-band modeling and deblending, i.e. separating individual astrophysical objects overlapping on the plane of the sky. We carefully design our algorithm to perform a linear resampling that explicitly accounts for the instrumental and optical response of the image acquisition process. This allows the \scarlet package to explicitely account for the resampling operator in the optimization process of modeling images.
    
    An alternative for pixel-level joint modeling and deblending at two different resolutions has been recently published in \cite{Arcelin2020}. In that work, the authors use a variational autoencoder that learns to recover the morphologies of galaxies by training them on blended images from multiple instruments. In contrast, we implement an explicit resampling operation that transforms a high-resolution image into a lower-resolution images to match any observation. By implementing this operation separately, can use with any type of parameterization for the high-resolution model and any type of inference.
    
    In this paper, we restrict ourselves to presenting the resampling method and validating that pixel-level joint survey deblending leads to better reconstruction of light profiles than single-resolution deblending. A more quantitative study of the gains from joint survey processing requires using metrics that are specific to the individual science cases of interest. This is outside of the scope of this paper. A follow-up publication will study the gains from using multiple resolution images specifically for object detection and its impact on deblending.
    
     The paper is organised as follows:  we introduce our resampling approach in Section \ref{Sect:resampling}. Section \ref{Sect:accelerate} discusses our implementation of a practical and faster resampling strategy in the \scarlet framework. We compare the runtime and accuracy of the resampling with the reference BG method implemented in \href{https://github.com/GalSim-developers/GalSim}{\faGithub \galsim}\footnote{https://github.com/GalSim-developers/GalSim} \citep{Rowe2015} in sections \ref{Sect:Timing} and \ref{Sect:accuracy} respectively. In section \ref{Sect:real_images}, we compare the results of joint deblending to individual survey deblending on simulated images of the {\it Euclid},  {\it Rubin}, and LSST surveys. We conclude in section \ref{Sect:conclusion}.

\section{Resampling and convolution of astronomical imaging}
\label{Sect:resampling}

    To maintain the flexibility of a pixelated galaxy model, combining surveys at different resolutions requires an explicit operation for resampling and convolution with the PSF. Both operations can be combined analytically.
    
\subsection{Image modeling}
\label{Sec:image-formation}

    Let the function $f:\mathbb{R}^2\to\mathbb{R}$ represent the intrinsic, continuous surface brightness of the sky. Acquisition of images by telescopes causes the function $f$ to be convolved by the PSF $p$ of the telescope. The telescope is equipped with a camera with $n_{pix}\times n_{pix}$ pixels, which then integrates $(p*f)$ over the pixel area and samples it at the pixel positions ${(\xk, \yl)}\ \forall {k_1,k_2 \in [\![0, n_{pix}]\!]^2}$ . Ignoring all notions of noise in the acquisition process for the purpose of the demonstration, an image can therefore be described as a set of discrete samples of a continuous function $I$ defined as:

\begin{equation}
    I(x,y) = (\rect^{2D}_h * (p * f)){(x,y)}, \: (x, y) \in \mathbb{R}^2, \label{eq:observation}
\end{equation}
    where $h$ is the linear size of a square pixel, i.e. $ h = \xk-x_{{k_1}-1} = \yl-y_{k_2-1} $. The function $\rect^{2D}_h$ is the 2-dimensional rectangle function defined by:

\begin{eqnarray}
\forall (x,y,t) \in \mathbb{R}^3, & \nonumber \\
\rect^{2D}_h(x, y) & =  \rect_h(x)\cdot \rect_h(y), \nonumber \\
\rect_h(t) & = 
  \begin{cases}
    0       & \quad \text{if } |t| > h/2, \nonumber\\
    1  & \quad \text{otherwise.} \nonumber
  \end{cases} \label{eq:imaging}
\end{eqnarray}
Convolution by $\rect^{2D}_h$, followed by sampling at the pixels center, is identical to the integration of the convolved surface brightness $(p*f)$ over the pixels surface.

\subsection{Whittaker-Shannon interpolation}
\label{Sect:sinc_interp}
    The Whittaker-Shannon interpolation states that a continuous, band-limited function, $f: \mathbb{R} \to \mathbb{R}$ can be recovered at \emph{any} point $x$ from an infinite number of sampled evaluations $f_k = f(x_k)$:  

\begin{equation}
f(x) = \sum_{k\in \mathbb{Z}} f_k\, \sinc\Bigg{(}\frac{x-x_{k}}{h}\Bigg{)},  \label{eq:whitt}
\end{equation}
    where $h$ is the sampling interval of points $x_{k}$, i.e. $x_{k}-x_{k-1} = h$ and $sinc$ is the normalized sinc function expressed as: 
    
    \begin{equation}
        sinc(x) = \frac{sin(\pi x)}{\pi x} \nonumber
    \end{equation}. 
    Equation \ref{eq:whitt} requires that $f$ is band-limited, which means that its Fourier transform $\hat{f}(\nu)$ has non-zero values only in the region $|\nu| \leq \frac{1}{2h}$.
    The Whittaker-Shannon interpolation formula trivially extends to two-dimensional functions:

\begin{equation}
f(x, y) = \sum_{(k_1,k_2)\in \mathbb{Z}^2} f_{{k_1},{k_2}}\, \sinc\Bigg{(}\frac{x-\xk}{h}\Bigg{)}\,\sinc\Bigg{(}\frac{y-\yl}{h}\Bigg{)}.  \label{eq:whitt2D}
\end{equation}

Since band-limited signals cannot be compact in real space, the summations in the previous equations have to be carried over the entire $\mathbb{Z}^2$ domain, which is not computationally tractable. In practice, summations are carried over a limited spatial domain, which is an approximation of the Whittaker-Shannon interpolation. With our target application being the modeling of individual astronomical objects (stars and galaxies), the approximation errors are limited by the fact that the flux of these objects is concentrated in a small region around their center. 

In the following equations we will abstract the summation domain such that it is implied that all summations are over $\mathbb{Z}$ or $\mathbb{Z}^2$, unless otherwise specified.

\subsection{Interpolation and convolution}
\label{Sec:conv-inter}

    Because the operations of convolution and resampling are computationally intensive, we seek a strategy to perform both as a single operation. We first consider the one-dimensional case of resampling and convolution of a function $f:\mathbb{R}\to\mathbb{R}$ sampled at $n$ positions $\xk$ spaced regularly by a length $h$ such that $\xk = k_1 h$. Our goal is to express the relation between these samples and the $n'$ samples of the function $f*p$ evaluated at  $n'$ linearly spaced positions $\Xp$ separated by $h'$, where we generally assume but not require $h'\neq h$. The convolution kernel $p$ is arbitrary, with $n_p$ samples at positions $X_{m_1} = m_1 h'$.  Throughout this calculation, we assume that $f$ and $p$ are band-limited for a grid with spacing $h$.
    Using equation \ref{eq:whitt}, we can represent function $f*p$ at any point $x\in [0, n h]$ as a function of the regular samples $f_{k_1}=f(\xk)$ and $p_{m_1}=p(\xii)$:
    
    \begin{equation}
        (f*p)(x) = \sum_{k_1} f_{k_1} \,\sinc\Bigg{(}\frac{x-\xk}{h}\Bigg{)} * \sum_{m_1} p_{m_1}\,\sinc\Bigg{(}\frac{x-\xii}{h}\Bigg{)}. \label{eq:conv_samp}
    \end{equation}
    With the Fourier transform of the sinc expressed as:
    
    \begin{equation}
        \widehat{\sinc\Bigg{(}\frac{x-\xk}{h}\Bigg{)}}(\nu) = h\times\rect_h(\nu) \,\mathrm{e}^{-i\nu \xk}, 
    \end{equation}
    where $\mathrm{e}^{-i\nu \xk}$ is the Fourier equivalent of a real shift by $x_k$, we can write the Fourier transform of equation \ref{eq:conv_samp}:
    \begin{eqnarray}
    \widehat{(f*p)}(\nu) = & h^2 & \sum_{k_1} f_{k_1}\, \rect_h(\nu) \,\mathrm{e}^{-i\nu \xk} \times \nonumber\\ 
    && \sum_{m_1} p_{m_1}\, \rect_h(\nu)\,\mathrm{e}^{-i\nu \xii} \nonumber\\
         = & h^2 & \rect_h(\nu) \sum_{k_1} f_{k_1} \sum_{m_1} p_{m_1}\,\mathrm{e}^{-i\nu (\xk+\xii)}.
    \end{eqnarray}
    Note here that the product of two identical top-hat functions with the same width gives the same top-hat function. As a result, when transforming back into direct space, a factor $h$ remains, such that
    \begin{equation}
        (f*p)(x) = h \sum_{k_1} f_{k_1} \sum_{m_1} p_{m_1}\, \sinc\Bigg{(}\frac{x-\xk-\xii}{h}\Bigg{)}. \label{eq:reconvampling}
     \end{equation}
     Equation \ref{eq:reconvampling} allows us to compute $(f*p)$ at any point $x\in [0, n h]$. In particular, we want to evaluate $(f*p)$ in the $n'$ positions $X_{l_1}$. Although equations \ref{eq:conv_samp} and \ref{eq:reconvampling} yield identical results, $n'$ samplings of equation \ref{eq:conv_samp} requires a convolution with complexity $\mathcal{O}(n'n')$ on top of the resampling of $f$, with complexity $\mathcal{O}(nn')$. Equation \ref{eq:reconvampling}, however, requires only one operation with complexity $\mathcal{O}(nn')$. In addition, the inner sum of equation \ref{eq:reconvampling} does not depend on $f$, which allows us to precompute it for iterative schemes like the optimization on \scarlet. 
     
     The two-dimensional case follows from equation \ref{eq:whitt2D} and the separability of the Fourier transform and the $\sinc$ function:
     \begin{equation}
    \begin{split}
         (f*p)(x, y) = h^2  &\sum_{k_1, k_2} f_{k_1,k_2} 
         \sum_{m_1,m_2} p_{m_1,m_2}\,\times\\ &\sinc\Bigg{(}\frac{x-\xk-\xii}{h}\Bigg{)}\,
         \sinc\Bigg{(}\frac{y-\yl-\yj}{h}\Bigg{)}.
    \end{split}
    \label{eq:reconvampling2D}
     \end{equation}
     By noticing that equation \ref{eq:reconvampling2D} defines a two-dimensional discrete convolution product by a shifted $\sinc$ function, it can be rewritten as:
     \begin{eqnarray}
         (f*p)(x,y)= h^2 \sum_{k_1,k_2} f_{k_1,k_2}\,(p_{m_1,m_2}*S_h(x-\xk, y-\yl)), \label{eq:compact2D}
     \end{eqnarray}
     where $S_h$ is the two-dimensional $\sinc$ kernel defined as
     \begin{equation}
        S_h(x,y) = \sinc\left(\frac{x}{h}\right)\,\sinc\left(\frac{y}{h}\right).
    \end{equation}     
     Equation \ref{eq:compact2D} defines a convolution product with three terms, $f$, $p$ and $S_h$, and it is linear in all three of them. 
     
\subsection{Resampling}
    We now combine the results of sections \ref{Sec:image-formation} and \ref{Sec:conv-inter}
    and apply the convolution scheme to images with different resolutions.
    Let us consider two different images $(I_1, I_2)$ of the same patch of the sky with two different resolutions, PSFs, and number of pixels, but in the same wavelength range. Following equation \ref{eq:observation}, these images are described by:
   \begin{equation}
   \begin{split}
        I_1(\xk,\yl) &= (rect^{2D}_{h_1}*f*p_1)(\xk, \yl) \\
        I_2(\Xp,\Yq) &= (rect^{2D}_{h_2}*f*p_2)(\Xp,\Yq),
    \end{split}
    \end{equation}
    where $(k_1, k_2)\in [\![0,n_1]\!]^2$ and  $(l_1,l_2)\in [\![0,n_2]\!]^2$.
    We use the following convention, with the hope that this will save the readers a few headaches: When indexing samples on two different grids, upper-case letters will refer to samples on the coarse grid $(\Xp, \Yq)$ and lower case letters will refer to samples on the finer grid $(\xk, \yl)$. That means $h_1\leq h_2$.

    The continuous functions $p_1$ and $p_2$ that describe each instrument's PSFs are generally not known, and are instead represented by their integrated and sampled values:
    \begin{equation}
    \begin{split}
        P_1(\xii, \yj) &= (\rect^{2D}_{h_1} * p_1)(\xii, \yj)\\
        P_2(X_{o_1}, Y_{o_2}) &= (\rect^{2D}_{h_2} * p_2)(X_{o_1}, Y_{o_2}).
    \end{split}
    \label{eq:psf}    
    \end{equation}
     We also assume that the Fourier support of $p_1$ is larger than that of $p_2$, which is usually satisfied when $h_1\leq h_2$ to ensure critical sampling of the PSFs.

    While the images $I_1$ and $I_2$ could describe multiple observations, we are interested in fitting models that can have higher resolution and sharper PSF than the observations. For non-parametric models, the model itself can be represented as an image. For the sake of this demonstration, where images are assumed noiseless, we choose to parameterize our fiducial model of the sky as the higher-resolution image $I_1$, which means that on top of the inescapable $\rect$ convolution, we allow it to maintain its own effective PSF $p_1$. This has the added benefit to maintain the band-limitedness of the model required for our interpolation scheme. 
    Due to the linearity of the convolution operation, and the ability to resample at any location, the combined operation can be expressed as a linear equation,
    \begin{equation}
        I_2 = R_2 I_1  \label{eq:A_2}
    \end{equation}
   With this in mind we can use equation \ref{eq:reconvampling2D} to build the operator $R_2$ in equation \ref{eq:A_2}. Because the model, $I_1$, is a sampling of the integrated surface brightness of the sky in pixels of size $h_1$, convolved by PSF $p_1$, operator $R_2$ has to account for three effects:
    \begin{itemize}
        \item the difference in PSF between $p_1$ and $p_2$,
        \item the resampling from positions $(\xk, \yl)$ to $(\Xp,\Yq)$,
        \item and the difference in integration surface between $I_1$ and $I_2$.
    \end{itemize}
    To address the first item, we need to compute the difference convolution kernel $P_d$ between $P_1$ and $P_2$.
    Using equation \ref{eq:psf}, we can express $P_d$ as a function of $p_1$ and $p_2$:
    \begin{equation}
        P_d = \mathcal{F}^{-1} \Bigg{(} \frac{\widehat{\rect^{2D}_{h_2} * p_2}}{\widehat{\rect^{2D}_{h_1} * p_1}} \Bigg{)},
    \end{equation}
    where $\mathcal{F}^{-1}$ stands for the inverse Fourier transform.
    This ratio is well-defined if the Fourier support of $p_1$ is larger than that of $p_2$, which we already required earlier. It covers the particular case of the vanishing PSF $p_1(\xii, \yj)=\delta(\xii, \yj)$ and that of the identity mapping where $\rect^{2D}_{h_1} * p_1=\rect^{2D}_{h_2} * p_2$ leads to $P_d = \delta$ . 
    To compute the difference kernel in practice, we need to have $P_1$ and $P_2$ sampled on the same grid. Because the grid of $I_1$ is chosen to be the reference grid and to prevent down-sampling of $P_1$, we build $P_{i_2}$, the interpolation of $P_2$ on $P_1$'s grid:
    \begin{equation}
        P^{(i)}_{2}(\xii,\yj) = \sum_{o_1, o_2}P_{2;o_1,o_2} \, S_{h_2}((\xii,\yj)-(X_{o_1}, Y_{o_2})), \label{eq:psf_interp}
    \end{equation}
    $P^{(i)}_{2}$ is the interpolation of $p_2$ integrated in windows of size $h_2$ but sampled in with pixel spacing $h_1$. We are now able to construct the difference kernel:
    \begin{equation}
        P_{d(m_1,m_2)} = \mathcal{F}^{-1}\Bigg{(}\frac{\hat{P}^{(i)}_{2}}{\hat{P}_{1}}\Bigg{)}(x_{m_1}, x_{m_2}),
    \end{equation}
    
    Using equation \ref{eq:compact2D}, we can interpolate the samples for $I_1(x_{k1},y_{k2})$ at the locations of $I_2$'s pixels by substituting $p$ for $P_d$ and $(x,y)$ for the set of $(\Xp, \Yq)$ values:
    \begin{eqnarray}
         & I_2(\Xp,\Yq)  \\
          = & h^2\sum_{k_1,k_2} (\rect^{2D}_{h_1}*f*p_1)_{k_1,k_2}(P_{d(m_1,m_2)}*S_h(\Xp-\xk, \Yq-\yl))\nonumber. \label{eq:I2model}
    \end{eqnarray}
    It follows that operator $R_2$ can be represented by a $n_1\times n_2$ matrix, whose elements are defined by the inner sum of the first line of equation \ref{eq:I2model}, and quantifies the contribution of model sample $I_{x_1,k_2}$ to image sample $I_2(\Xp,\Yq)$. This expression holds because the difference kernel carries the ratio of pixel integrations $\widehat{rect^{2D}_{h_2}}/\widehat{rect^{2D}_{h_1}}$ between images of $p_1$ and $p_2$. 
    
\section{Accelerated resampling}
\label{Sect:accelerate}

    According to equations \ref{eq:reconvampling2D} and \ref{eq:I2model}, resampling and convolving an image requires one to compute the matrix $R_2$ of size $(n_1^2\times n_2^2)$ and to perform a matrix multiplication by $R_2$ every time the resampling and convolution is applied. We remind the reader that the target application for this scheme is iterative model fitting with \scarlet, which implements a gradient descent optimisation. The goal of the acceleration presented here is therefore to minimise the time spent on computations at each iteration.
    We seek to reformulate equation \ref{eq:reconvampling2D} to accelerate these operations, which should be possible due to the separability of the $\sinc$ function.

    \subsection{The case of non-rotated grids}
    \label{sec:fast_aligned}
    Let a high resolution image $I_1$ be of size $n_{1x} \times n_{1y}$ and its low resolution counter-part $I_2$ with size $n_{2x} \times n_{2y}$ verifying equation \ref{eq:I2model}. First, we notice that there are in total $n_{1x}$ different values of pixel coordinates $(\xk,\yl)$ for $I_1$ on its x-axis and $n_{1y}$ different values on its y-axis. Similarly, in the case where the grids of $I_2$ and $I_1$ are aligned, there are only $n_{2x}$ and $n_{2y}$ different values for the pixel coordinates $(\Xp,\Yq)$ of $I_2$ along its x and y-axes respectively. As a result, it is only necessary to compute $n_{2x}\times n_{1x} +n_{2y}\times n_{1y}$ evaluations of 1-dimensional $\sinc$ kernels to determine the $n_{2y}\times n_{2x}\times n_{1y}\times n_{1x}$ matrix $R$. This realization allows a faster computation of $R$, but it is possible to achieve even faster resampling by separating the $\sinc$ kernel in its 1-dimensional components and factoring it such that equation \ref{eq:I2model} applied to $I_2$ and $I_1$ becomes:
    \begin{equation}
    \begin{split}
        I_2(\Xp,\Yq) = h_1^2 \sum_{k_1,k_2} &\sum_{m_1}  I_1(xk_1,xk_2)\, \sinc\Bigg{(}\frac{\Xp-\xk-\xii}{h_1}  \Bigg{)} \\
        &\sum_{m_2} P_{d;m_1,m_2}\,\sinc\Bigg{(}\frac{\Yq-\yl-\yj}{h_1}\Bigg{)}.
    \end{split}
    \label{eq:sinc_sep}
    \end{equation}
The second line defines $n_{2y}$ one-dimensional convolution products by $\sinc$ functions shifted by $\Yq$. The first line also contains one-dimensional convolution products along the x-axis of image $I_1$, by $\sinc$ kernels shifted by $\Xp$. The first two summations define a matrix product between the results of the two one dimensional convolutions. A more compact expression for equation \ref{eq:sinc_sep} is therefore:
    
    \begin{equation}\small
        I_2(\Xp, \Yq) = h_1^2  \sum_{\yl} \sum_{\xii} (I_1\underset{1}{*}s_{h1})(\Xp-\xii, \yl) (P_d\underset{1}{*}s_{h1})(\xii, \Yq-\yl), \label{eq:compact_sinc_sep}
    \end{equation}{}
    where symbol $\underset{1}{*}$ stands for the one-dimensional convolution and $s_h$ is the 1-dimensional $sinc$ function defined as $s_h(x) = sinc(\frac{x}{h})$. Equation \ref{eq:compact_sinc_sep} defines two convolution products. However, given that samples $(\Xp, \Yq)$ and $(\xk,\yl)$ are not on the same grid, fast algorithms for convolutions do not apply.
    
    By evaluating these convolutions and shifts in the Fourier domain, using the rectangle function as the analytic Fourier transform of the $\sinc$, we find that this scheme allows us to perform resampling ten times faster than when using the full resampling matrix $R$ on typical runs with a high resolution image of size ($250\times 250$) and its low resolution counter-part with size ($50\times 50$). Also, the setup time spent in computing matrix $R$ on typical runs is divided by a factor $\sim 100$ by only computing one-dimensional convolutions and shifts on $P_d$.
    
    Here, we chose to associate the $\sinc$ with arguments along the $x$-axis with function $I_1$ and the $\sinc$ with arguments along the $y$-axis to $P_d$ in equation \ref{eq:sinc_sep}, however we could just as easily have inverted this choice. In cases where images have rectangular shapes (meaning $n_{1x}\neq n_{1y}$), this freedom allows us to optimize the number of operations by choosing to convolve $I_1$ by a $\sinc$ along its smallest dimension. 
    
    \subsection{Generalization to rotated frames}
    \label{sec:fast_rotated}
\begin{figure*}[ht!]
        \centering
        \includegraphics[scale = 0.5]{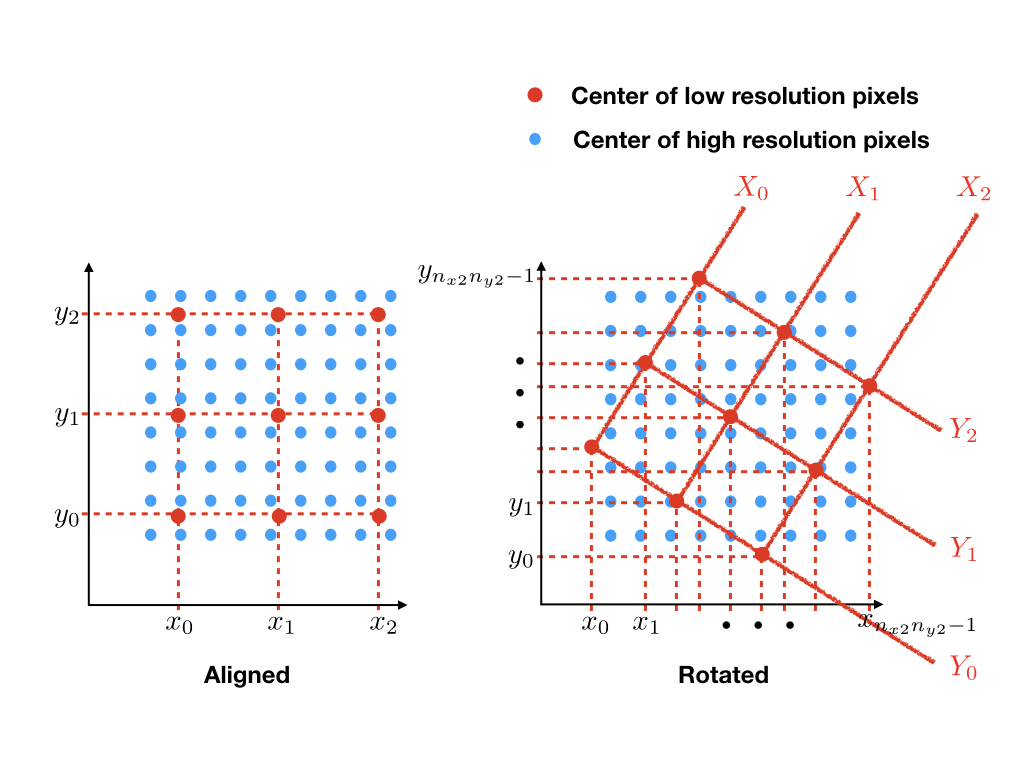}
        \caption{Illustration of the pixel coordinates distribution in the aligned and rotated cases. The blues points represent the positions of the centers of high resolution pixels, while red points give the positions of figurative centers of low resolution pixels. There are $n_{2x}\times n_{2y}$ red points in each panel. Projected on the $x$ and $y$ axes in the aligned panels are the $n_{2x} + n_{2y}$ values needed to identify these red points. On the rotated panel, the red points are described by $n_{2x} + n_{2y}$ values projected on a rotated frame depicted by the capital letters $X, Y$.}
        \label{fig:rotation}
    \end{figure*}

    Two images of the same patch of sky might happen to be aligned, but in general this assumption is not true.
    The technique we just described is a special case of a more general formalism that we introduce in this section. 
    In the case where images are rotated with respect to one another, the number of different values $\Xp$ and $\Yq$ can take is not limited to $n_{2x}$ and $n_{2y}$ but can be as large as $n_{2x}\times n_{2y}$ for each of them. This is illustrated in Figure \ref{fig:rotation}, where the rotated images on the right hand-side show pixel positions $\Xp$ and $\Yq$ that have nine different values each, while aligned images as represented on the left have three different values each.
    
    In the rotated case, using the splitting outlined in section \ref{sec:fast_aligned} leads to large matrix multiplications and is actually detrimental to the speed of the operation, making it a worse formulation than the naive equation \ref{eq:reconvampling2D}.  
    
    We can find a more general formulation for the factorization of equation \ref{eq:reconvampling2D} by:
    
    \begin{itemize}
        \item firstly, transforming coordinates $(\Xp, \Yq)$ to an un-rotated frame where all pixel coordinates $\Xp$ and $\Yq$ can be expressed by $n_{2x}$ and $n_{2y}$ different values respectively.    
        \item secondly, writing equation \ref{eq:reconvampling2D} as the three-factor convolution product that it is, we can use the associative property to split and redistribute the arguments of the discrete convolution along carefully selected axes.
    \end{itemize}

    \subsubsection{Frame rotation}
    
    Because $I_2$ is acquired on a regular grid, there should exist a frame in which its coordinates take a minimal number of different values $(X_m, Y_n)$. This frame is trivially given by applying to the coordinates $(\Xp,\Yq)$ of $I_2$, the rotation matrix which angle matches the angle $\theta$ between both images such that:
    
    \begin{eqnarray}
    \Xp & = & X_m \cos{\theta} + Y_n \sin{\theta}, \nonumber \\
    \Yq & = & Y_n \cos{\theta} - X_m \sin{\theta}. \label{eq:rotation}
    \end{eqnarray}
    
    \subsubsection{Convolution associativity}
    
    By realizing that equation \ref{eq:reconvampling2D} is a three-factor discrete convolution that generates a result sampled on a different grid than two of its factors, we can write $I_2$ as:
    \begin{equation}
        I_2 = I_1*P_d*S_h
    \end{equation}
    
    By associativity and commutativity of the convolution product and by substituting equation \ref{eq:rotation}, we can rewrite equation \ref{eq:reconvampling2D} as:
    
    \begin{eqnarray}
         I_2(\Xp,\Yq)& = & h^2  \sum_{k_1,k_2} I_1(x_{k1},x_{k2})  \sum_{m_1,m_2} P_{d;m_1,m_2}\, S^{m,n}   \nonumber \\
         &=& h^2  \sum_{k_1,k_2,m_2} I^n(\xk, \yl)\,\sinc\Bigg{(}\frac{\yl+\yj}{h}\Bigg{)} \times \nonumber \\  && \sum_{m_1} P^m_d(\xii,\yj)\,  \sinc\Bigg{(}\frac{\xk+\xii}{h}\Bigg{)},
         \label{eq:fast_reconvampling}
    \end{eqnarray}
    with:
    \begin{eqnarray}
        S^{m,n} & = & \sinc\Bigg{(}\frac{X_m \cos{\theta} + Y_n \sin{\theta}-\xk-\xii}{h}\Bigg{)} \nonumber \\
        && \sinc\Bigg{(}\frac{Y_n \cos{\theta} - X_m \sin{\theta}-\yl-\yj}{h}\Bigg{)} \nonumber \\
        F^n(\xk, \yl) & = & I_1(\xk - \frac{Y_n}{h} \sin{\theta}, \yl - \frac{Y_n}{h} \cos{\theta}) \nonumber \\
        P^m_d(\xii,\yj) & = & P_d(\xii-\frac{X_m}{h} \cos{\theta},\yj+\frac{X_m}{h} \sin{\theta}). \nonumber
    \end{eqnarray}
    In the last equality of equation \ref{eq:fast_reconvampling}, the first term in $I_1$ is computed by shifting image $I_1$ in the Fourier domain in $n_{2y}$ different directions $(Y_n \sin{\theta}, Y_n \cos{\theta})$. The remainder is obtained by performing $n_{2x}$ shifts of the result of the convolution between $P_d$ and $S_h$ by vectors with components $(X_m \cos{\theta}, - X_m \sin{\theta})$. 
    
    Once again, we chose to pick the terms in $Y_n$ for the shifts in $I_1$, but an equivalent result would be achieved by having $I_1$ shifted by $(X_m \cos{\theta}, - X_m \sin{\theta})$ and $P_d*S_h$ shifted by $(Y_n \sin{\theta},\\ Y_n \cos{\theta})$. While it might seem unsettling to see terms in $Y_n \sin{\theta}$ associated with the $x$-axis of $I_1$, the reader should keep in mind that this term comes from the decomposition of $\xii$. It is also important that both axes be shifted by values depending on the same ``unrotated'' variable, ($Y_n$ for the shift in $I_1$, in equation \ref{eq:fast_reconvampling}), otherwise there exists as many combinations of $(X_m, Y_n)$ as there are pixels in $I_2$ and the computational gain is lost.
    
    \subsubsection{Link with the aligned case and extensions}
    
    The acceleration from section \ref{sec:fast_aligned} qualifies easily as a particular case of the scheme described in section \ref{sec:fast_rotated}. The main semantic difference between these formulations is that, in the former we used the separability of the $\sinc$ function to perform the dimensionality splitting, while in the latter, we reformulated the discrete convolution product. Actually, the result from equation \ref{eq:compact_sinc_sep} can be obtained by using the same convolution reformulation and then factoring by the one-dimensional $\sinc$ factors. Alternatively, in equation \ref{eq:fast_reconvampling}, it is possible to factorize one of the one dimensional $\sinc$ factors and have $I_1$ convolved by this $\sinc$. 
    
    By setting the value of $\theta$ to zero, equations \ref{eq:compact_sinc_sep} and \ref{eq:fast_reconvampling} are equivalent, if not for the choice of the axis along which the shift is operated. We choose here to show both approaches of $\sinc$ factorization and convolution reformulation for two reasons. First, we think that the factorization formulation makes an easier introduction to the notion of dimension splitting for acceleration and would thus ease the understanding of subsection \ref{sec:fast_rotated}. Second, because this separation reflects the content of our implementation in \scarlet. While both formulations are strictly equivalent and lead to the same result when $\theta=0$, the convolution reformulation in its implementation suggests 2-dimensional Fourier transforms with 2-dimensional shifts, one of them being of magnitude 0. This requires twice as many operations than with the factorization formulation where the Fourier transform is applied along one dimension only and the number of multiplications for the shift is doubled. 
    
    The dimensional splitting we deployed in this section to perform fast convolutions onto rotated frames can actually be generalized to any stationary, separable transform on the grid. Indeed, the matrix rotation used in equation \ref{eq:rotation} can easily be replaced by a matrix with arbitrary coefficients as long as they are independent on the grid's coordinates, or at least as long as they don't introduce cross terms that depend on both $X_m$ and $Y_n$. This extends to polynomial transformations with the condition that they contain no rectangular terms in $X_m^a \times Y_n^b$ and can be applied to conventional convolutions, including interpolations, and does not require the separability of the kernel.

    \hide{
    \subsubsection{\red{The PSF difference kernel is the perfect interpolator}}
    
    Equation \ref{eq:fast_reconvampling} contains integer terms in the $sinc$ functions, representing pixel positions on the high resolution grid normalized by the size of the high resolution pixels. As a result, these terms vanish for $\yl \neq -\yj$ and $\xk \neq -\xii$ and are equal to one otherwise. As a result, equation \ref{eq:fast_reconvampling} can be rewritten as:
    
    \begin{eqnarray}
        (f*p)(\Xp,\Yq) = & h^2 \sum_{\xk,\yl}  I^n(\xk, \yl) P^m_d(-\xk,-\yl). \label{eq:Compact_resampling}
    \end{eqnarray}
    
    By having the terms in $Y_n$ put back as arguments of $P_d$, we retrieve the general expression for the interpolation formula:
    
    \begin{equation}
        f(X,Y) = \sum_{x, y} f(x,y) K(x-X, y-Y), \label{eq:interp}
    \end{equation}
    
    where the interpolation kernel $K$ is replaced by the PSF difference kernel $P_d$. In other words, the sinc interpolation and convolution strategy we've implemented boils down to an interpolation where the interpolating kernel is the oversampled difference kernel. 
    }

\section{Speed and complexity}
\label{Sect:Timing}
    To test the performance of our algorithm, we implemented it in the \scarlet framework, and compare the run-time and the interpolation accuracy with a thoroughly tested reference implementation for resampling: \galsim. 

    Implementing the formulation described above requires several operations, whose computational complexity depends on the number of pixels in the high- and low-resolution frames. In this section we compute the complexity of these operations for square images with $M\times M$ pixels in the low-resolution channel and $N\times N$ in the high-resolution channel. Given that this algorithm is meant for an iterative framework, we make a difference between setup operations that are performed once, and running operations that have to be performed every time the model changes.
    
    Setup operations comprise:
        \begin{itemize}
            \item Fourier transform of the low- and high-resolutions PSFs: $\mathcal{O}(M^2\log(M)) + \mathcal{O}(N^2\log(N))$
            \item $N$ shifts of the low-resolution PSF in the Fourier domain (multiplication by a complex exponential along one direction: $\mathcal{O}(NM^2)$
            \item inverse Fourier transforms of the shifted PSFs: $\mathcal{O}(NM^2\log(M))$,
            \item matrix multiplications by the $\sinc$ kernel along each axis: $\mathcal{O}(M^2N^2)$,
            \item transform of the resulting interpolated PSF at high-resolution: $\mathcal{O}(N^2\log(N))$,
            \item division by the high-resolution PSF to obtain $P_d$: $\mathcal{O}(N^2)$ 
            \item apply $M$ shifts of $P_d$ in the Fourier domain along the coordinates of the low resolution image: $\mathcal{O}(MN^2)$
        \end{itemize}
    Run operations execute these steps every iteration:
        \begin{itemize}
            \item Fourier transform the high-resolution image along one axis: $\mathcal{O}(N^2\log(N))$, 
            \item apply $M$ shifts in the Fourier domain: $\mathcal{O}(MN^2)$
            \item apply inverse FFT to each shifted image along one axis: $\mathcal{O}(MN^2\log(N))$ 
            \item multiply by $P_d$ to obtain the low-resolution image $\mathcal{O}(M^2N^2)$
        \end{itemize}
    the total cost of which is therefore
    \begin{equation}
        \mathcal{C}_s = \mathcal{O}\Big{(}N^2(M+1) \big{(}\log(N) + M\big{)}\Big{)} \label{eq:scarlet_time}
    \end{equation}
    The complexity is dominated by the term in $M^2N^2$ which is expected for the Whittaker-Shannon interpolation. While this scheme does not reduce the complexity, we implement it as a matrix multiplication between two matrices with a total number of elements $MN^2$ each. A classical interpolation, where the kernel is applied without separation, would amount to multiplying a matrix with $M^2N^2$ elements by a $N^2$ elements vector. While these two strategies require the same number of operations, our approach has a smaller memory footprint as we do not need to store a large $M^2N^2$ elements matrix. In practice, this leads to a 10-fold speed up on a personal laptop. 

    To assess the performance of our resampling scheme, we compare it against the resampling scheme described in BG and implemented in the \galsim package as the default scheme. In their approach, BG use an approximation of the Whittaker-Shannon interpolation by a quintic kernel with only 6 elements along each dimension, applied in Fourier space, and rely on 4-fold padding of the input image to limit the interpolation errors. Convolution is then performed with \galsim's default setting in a separate operation at the resolution of the resampled image in the Fourier domain.
    For comparison, in the resampling scheme from BG, the authors report a complexity of:

    \begin{equation}
        \mathcal{C}_g = \mathcal{O}\Big{(}N^2\big{(}k^2 + s^2 \log(sN)\big{)}+M^2\big{(}3+\log(M)\big{)}\Big{)}, \label{eq:galsim_time}
    \end{equation}
    where $k=6$ is the size of the preferred quintic interpolation kernel and $s = 4$ is the default 4-fold padding necessary to prevent the formation of ghosts in the image.   
    Comparing equations \ref{eq:scarlet_time} and \ref{eq:galsim_time} suggests a theoretical factor of a few hundreds between the run-times of our implementation and that of \galsim's default for reasonable values of $M$ and $N$.


    We compare the actual run-time, without setup, of our implementation, $t_s$, with the run-time for the BG scheme in \galsim's default, $t_g$. To do so, we generate images at 5 different resolutions to emulate observations with current or upcoming telescopes. The PSFs used in these simulations are Gaussian profiles. Pixel scales and PSF sizes (standard deviations) are given in Table \ref{tab:resolutions}.
    
    \begin{table}[]
    \centering
        \begin{tabular}{c||cc}
             & $p$ & $\sigma_\text{psf}$  \\
            \hline
             HST & 0.06 & 0.074 \\
             EUCLID & 0.1 & 0.16 \\
             ROMAN & 0.11 & 0.186 \\
             HSC & 0.167 & 0.62 \\
             RUBIN & 0.2 & 0.7
        \end{tabular}
        \caption{Pixel scales, $p$ and PSF sizes $\sigma_\text{psf}$ used in the simulations. All values are given in arcseconds.}
        \label{tab:resolutions}
    \end{table}
    
    Here, each observation that has a larger pixel size than another, also has a larger PSF, meaning that comparing resolutions is consistent. For each telescope, except {\it Rubin}, we generate seven high-resolution images with 20, 30, 50, 70, 100, 150 and 200 pixels on a side. For a given observation, we then resample the image on a grid that spans the same area on the plane of the sky but at the resolution of all telescopes that have a larger pixel size using either our new or the \galsim resampling schemes. 
    
\begin{figure}[t]
        \includegraphics[width = \linewidth]{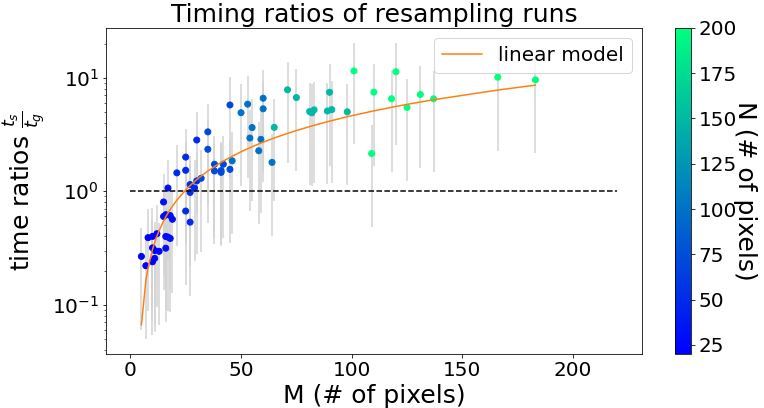}
        \caption{Time ratio, $\frac{t_s}{t_g}$ between \scarlet and \galsim's default implementations of resamplings in blue as a function of the number of samples in the low resolution channel for all couples of observations. The orange line is the best fit to the \scarlet comparison with a linear model. The color scale gives the number of samples in the low resolution channel for each point.}
        \label{fig:time_ratios(M)}
    \end{figure}

The results of this comparison are shown in Figure \ref{fig:time_ratios(M)}, where we see that the dependence between both strategies is in fact well described by a linear model:
\begin{equation}
    t_s / t_g \approx -0.17 + 0.048 M.
\end{equation}
The relative complexity does appear dominated by a term $\propto M$ instead of $M^2$ as suggested by equations \ref{eq:scarlet_time} and \ref{eq:galsim_time}.
    This is most likely explained by internals of the \galsim package that we have not accounted for, but we cannot provide a compelling explanation at this stage. While our algorithm is still prohibitively slow for large images compared to \galsim, it is reassuring that its performance in a python implementation is within a factor of a few of the current state-of-the-art algorithm.
    
\begin{figure}[t]
        \includegraphics[width = \linewidth]{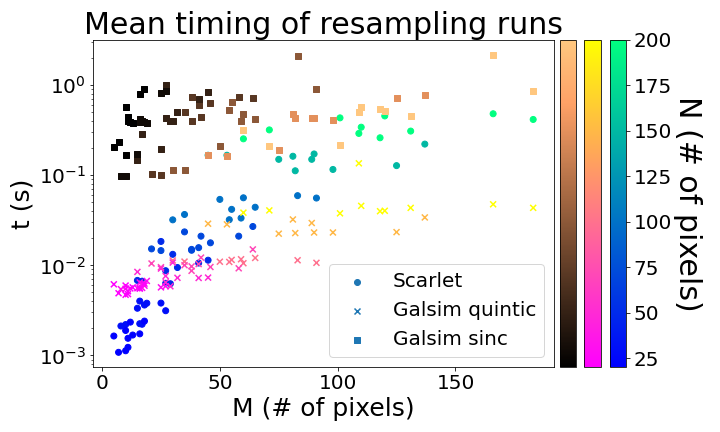}
        \caption{Average duration of \scarlet and \galsim resamplings over 1000 realisations.}
        \label{fig:absolute_timings}
    \end{figure}

    For reference we also show in Figure \ref{fig:absolute_timings} the average timings of \scarlet and \galsim's sinc and quintic resamplings. We observe that our implementation is faster than \galsim's sinc implementation for scenes where the high resolution image is smaller than 200 pixels. For larger images timing seems comparable. For future works, it might be of interest to implement approximate interpolation schemes to reduce computation times. For separable kernels, the acceleration presented in Section \ref{Sect:accelerate} would still provide benefits, thus enabling larger kernels that would mitigate approximation errors to remain computationally tractable.
    
    \hide{
    \begin{figure*}[ht!]
        \centering
        \begin{tabular}{cc}
        \includegraphics[scale = 0.35]{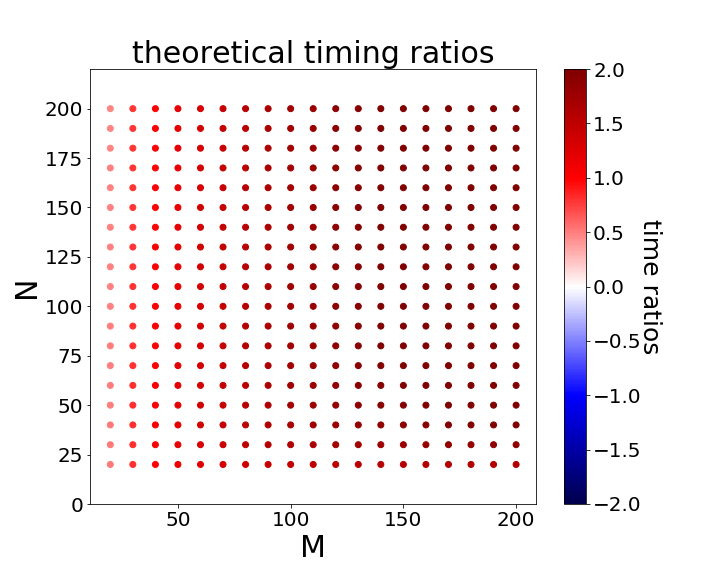} &
        \includegraphics[scale = 0.35]{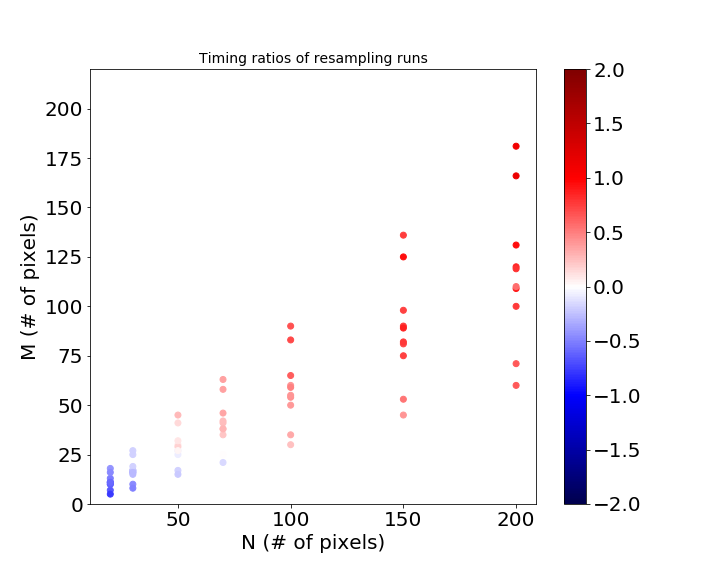}
        \end{tabular}
        \caption{{\it left handside}: theoretical time ratios between \scarlet and \galsim's default: $\frac{ \mathcal{C}_s}{ \mathcal{C}_g}$. {\it Right handside}: measured time ratios, $\frac{t_s}{t_g}$. Time ratios as reported in the color bars are in logarithmic scale.}
        \label{fig:theory_comparison}
    \end{figure*}
    }

    All the figures presented in this section can be reproduced from notebooks \href{https://github.com/herjy/Multi_resolution_comparison/blob/master/Timing_test/Timing_comparison.ipynb}{\faGithub Timing\_test}\footnote{https://github.com/herjy/Multi\_resolution\_comparison/ \\ blob/master/Timing\_test/Timing\_comparison.ipynb} and \href{https://github.com/herjy/Multi_resolution_comparison/blob/master/Timing_test/Time\%20fit.ipynb}{\faGithub Timing\_fit}\footnote{https://github.com/herjy/Multi\_resolution\_comparison/ \\ blob/master/Timing\_test/Time\%20fit.ipynb}. 

\section{Resampling accuracy}
    \label{Sect:accuracy}
    To test the quality of the reconstructions enabled by our implementation, we compare it with reconstructions from \galsim on simulated images.
    We use \galsim to simulate a set of 10,000 images of real galaxy profiles from the COSMOS sample \citep{Mandelbaum2012, Mandelbaum2014}, at the resolution of the emulated surveys from Table \ref{tab:resolutions}. Each of these images are drawn on a $60\times 60$ pixels grid, chosen to represent extended complex galaxy profiles that can be resampled in a timely fashion. These images serve as low resolution images, which are the ground truth for our reconstructions. For each of them, we use \galsim to produce images of the galaxies with the PSF and samplings of higher resolution surveys. 
    
    We then use our method and \galsim's default quintic interpolation to resample the high-resolution images to their low-resolution version and compare the result with the ground truth. While both ground truth images and \galsim resampled images are produced using the same software, the way they are produced is fundamentally different. Ground truth images are drawn directly from \galsim's {\tt GSObjects}. Resampled images are interpolated from the pixels of the high resolution images. For reference, we also reconstruct low resolution images using galsim's implementation of the sinc interpolation.
    
    For each reconstructed low resolution image, we compute the source distortion ratio of the reconstruction defined as:
    \begin{equation}
        SDR(\tilde{X}) = 10\log_{10}\Big{(}\frac{||X||}{||\tilde{X} - X||} \Big{)},
    \end{equation}
    where $\tilde{X}$ is the reconstructed low resolution image, $X$ is the true low resolution image and $||.||$ is the Euclidean 2-norm. The higher the SDR, the more accurate a reconstruction is.
    
\begin{figure*}[h!]
        \centering
        \includegraphics[width=\linewidth]{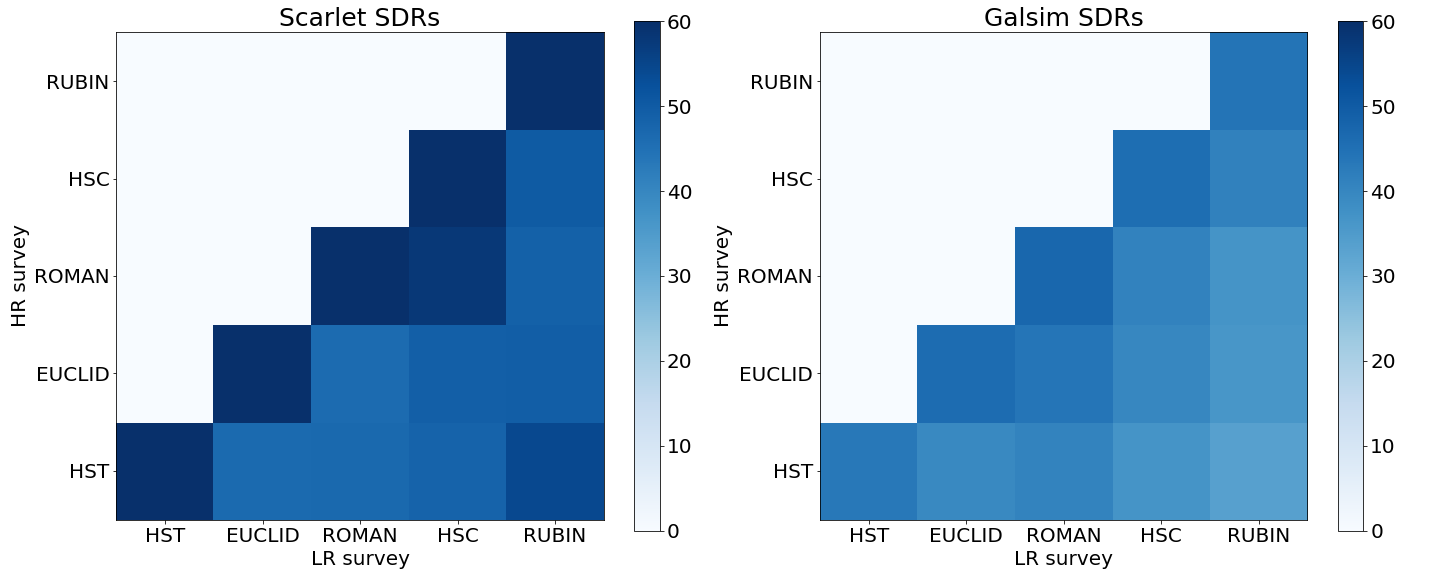}
        \caption{Mean SDRs of the \scarlet (left) and \galsim (right) reconstructions for each couple of resolutions. The y-axis shows the high resolution observation from which the reconstruction is built. The x-axis shows the target resolution. Initials HR and LR stand for high resolution and low resolution respectively.}
        \label{fig:SDR}
    \end{figure*}

\begin{figure*}[h!]
        \centering
        \includegraphics[width=\linewidth]{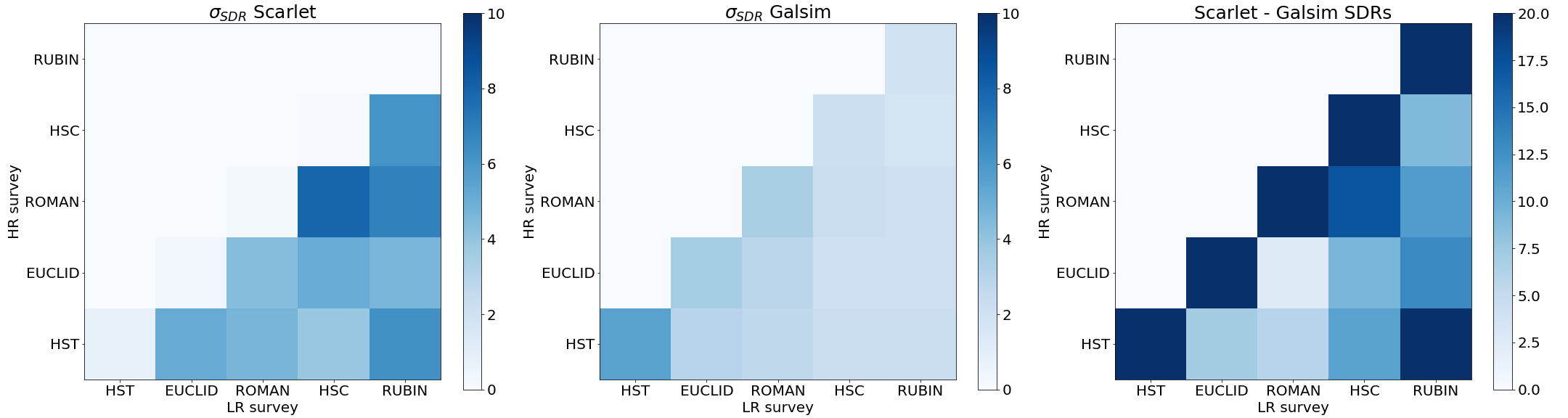}
        \caption{Standard deviation of the SDRs of the \scarlet (left) and \galsim's default (middle) reconstructions. Right hand-side panel shows the difference between the mean \scarlet and \galsim SDRs as shown in Figure \ref{fig:SDR}.}
        \label{fig:SDR_diff}
    \end{figure*}

\begin{figure*}[h!]
        \centering
        \includegraphics[width=\linewidth]{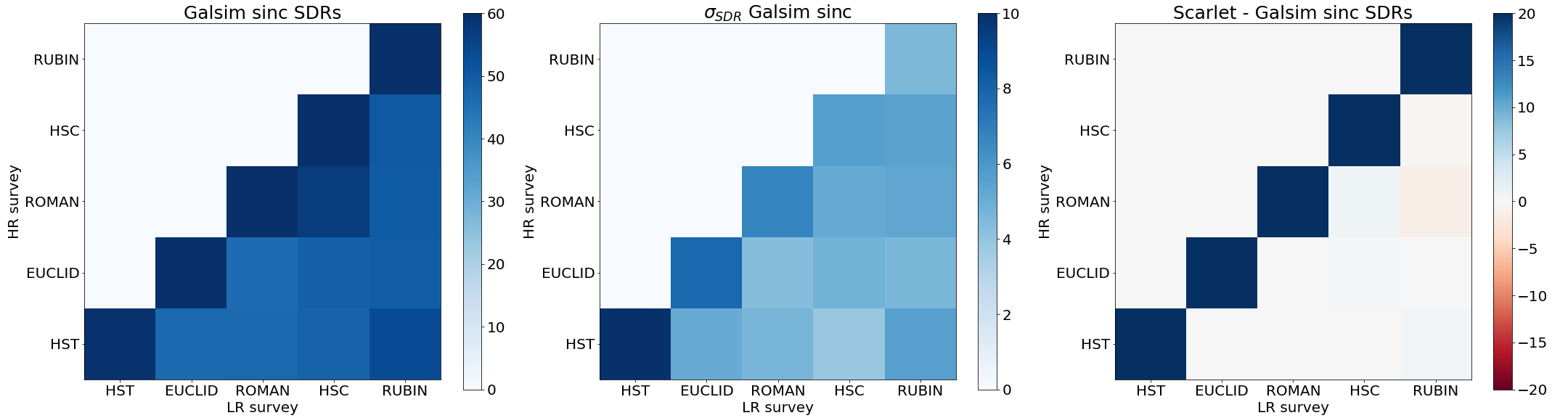}
        \caption{Mean and standard deviation of the SDR from \galsim's sinc interpolation (right and middle panels respectively). Right hand-side panel shows the difference between the mean \scarlet and \galsim sinc interpolation SDRs.}
        \label{fig:SDR_sinc}
    \end{figure*}

\begin{figure*}[h!]
        \centering
        \begin{tabular}{ccc}
        HST to {\it Rubin} & {\it Euclid} to {\it Euclid} & {\it Euclid} to {\it Rubin}\\
        \includegraphics[clip, trim = 0cm 0cm 0cm 2cm, scale = 0.23]{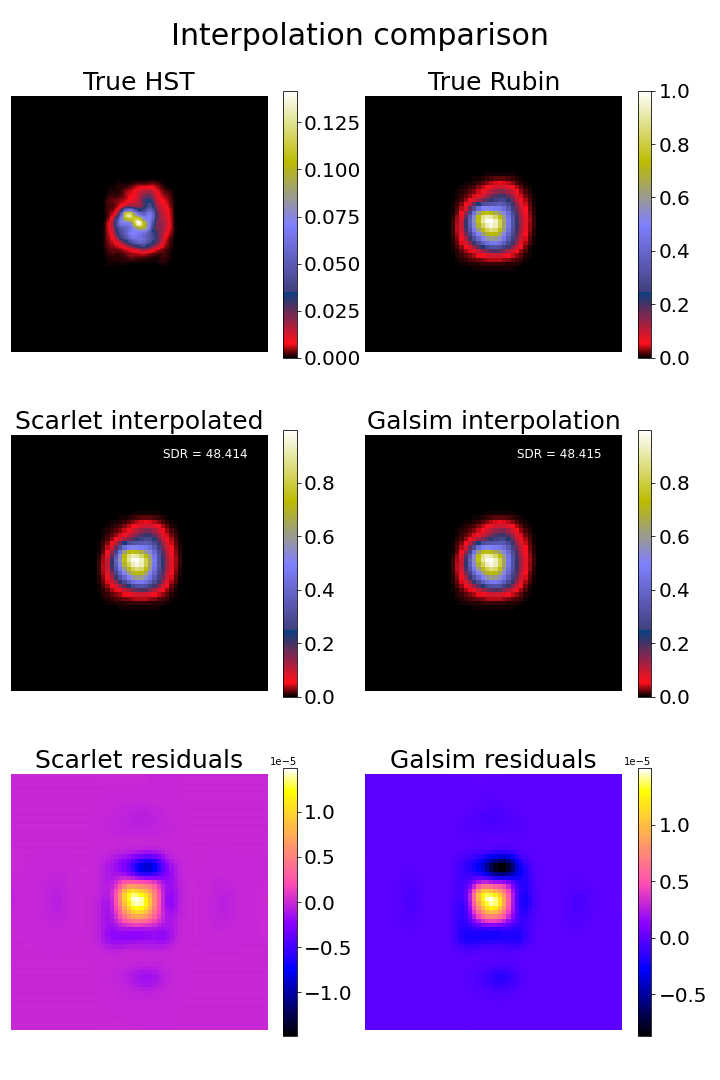} &
        \includegraphics[clip, trim = 0cm 0cm 0cm 2cm, scale = 0.23]{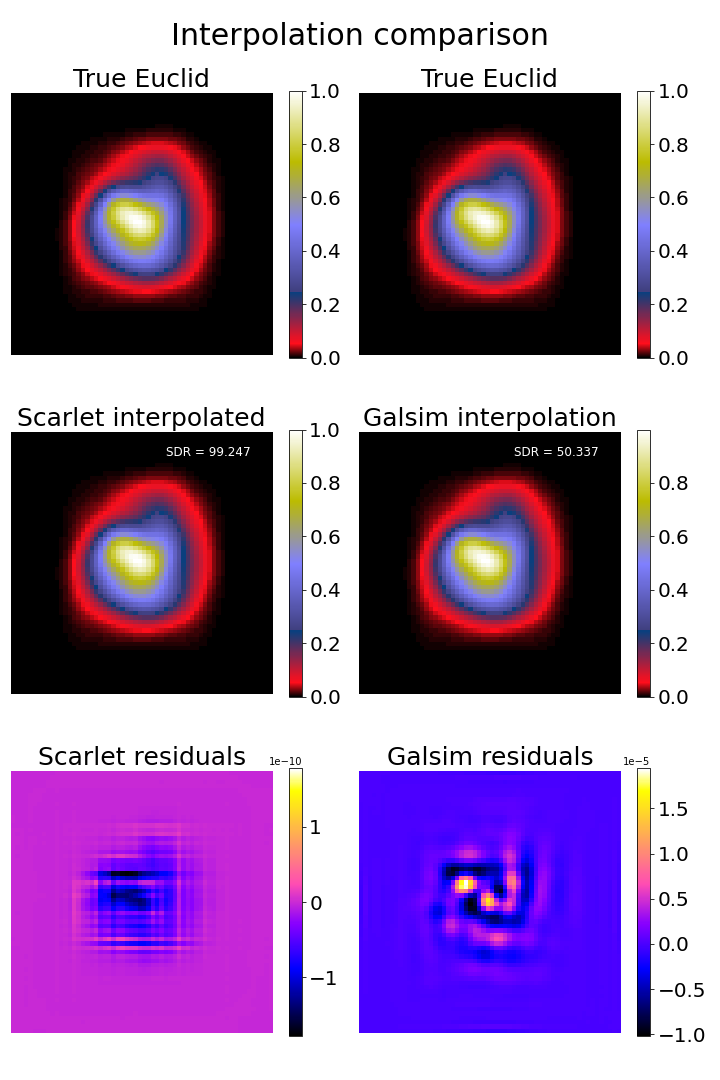} &
        \includegraphics[clip, trim = 0cm 0cm 0cm 2cm, scale = 0.23]{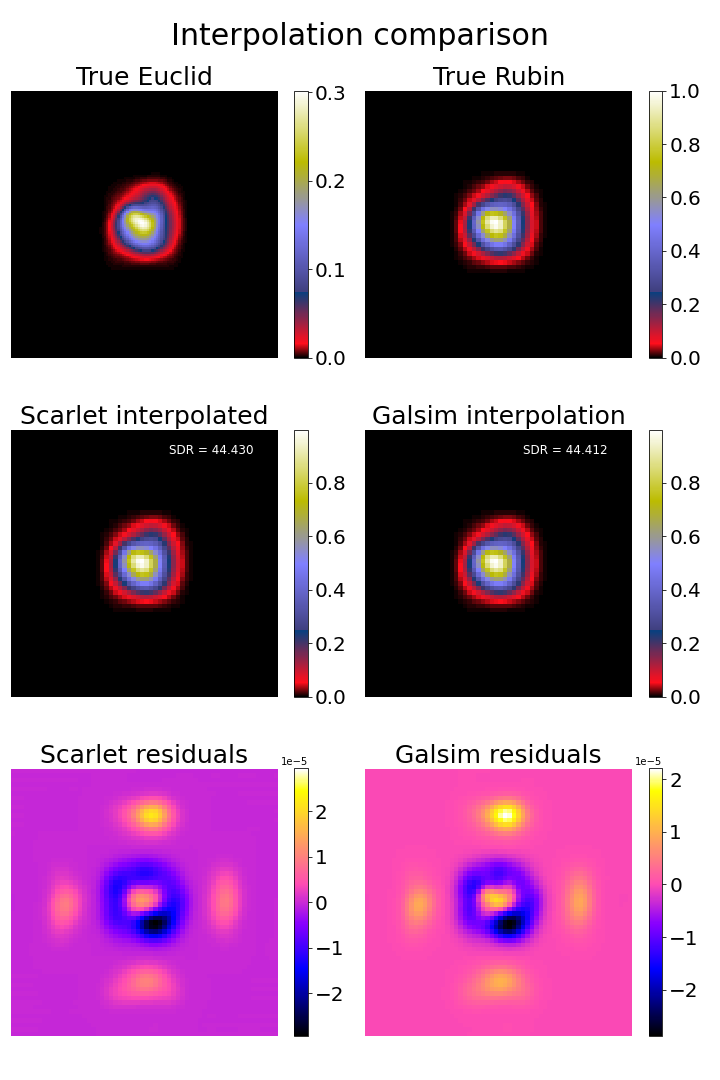}
        \end{tabular}
        \caption{Visual comparison of resampled images using \scarlet and \galsim. The first line of each column shows the the ground truth as generated by \galsim for the high resolution image (to the left of each column) and the low resolution image (to the right of each column). The second row shows the interpolation of the low resolution image from the high resolution data using \scarlet (left) and \galsim (right). The last row shows the difference between the \scarlet and \galsim reconstruction to the ground truth low resolution image.}
        \label{fig:visual}
    \end{figure*}

    We show the SDRs for the \scarlet and the \galsim reconstructions averaged over the different galaxy realizations in Figures \ref{fig:SDR} and \ref{fig:SDR_diff}. The average SDR difference is always positive, indicating that the \scarlet's resampling accuracy exceeds \galsim's default resampling in every pair of surveys. The most obvious differences are seen in the diagonal elements. They correspond to a resampling of a survey image on itself. In this case, \scarlet reproduces the identity operator but \galsim does not. As the ratio between pixel sizes increases, the \galsim's default scheme tends to yield worse SDRs on average. SDRs for \scarlet seem rather uniformly high without any noticeable dependence between surveys. This is to be expected as the reconstruction method in \galsim is an approximation, while \scarlet's reconstructions use a perfect interpolation scheme in the limit of band-limited signals. Results from \galsim's sinc resamplings are shown in figure \ref{fig:SDR_sinc} and show similar quality of reconstruction. Only along the diagonal axis that represents the identity mapping do we notice significantly different results, but even there, \galsim's sinc interpolation shows far better SDR than in the default case.  
    Our findings are confirmed by a visual inspection of the resampling residuals in Figure \ref{fig:visual}.
    
\begin{figure}[h!]
        \centering
        {\it Rubin} to {\it Rubin} \\
        \includegraphics[clip, trim = 2cm 2cm 2cm 4cm, scale = 0.4]{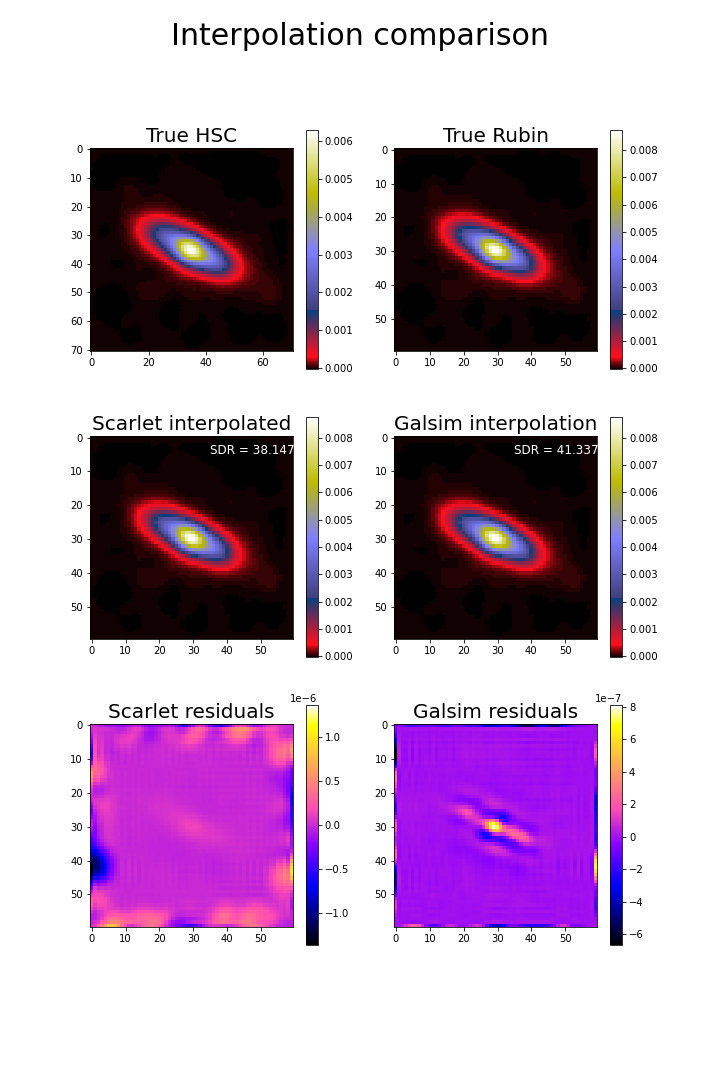}
        \caption{Same figure as in \ref{fig:visual} in a case where edge effects  lead to a \scarlet SDR lower the the \galsim SDR.}
        \label{fig:visual_edge}
    \end{figure}

    We notice that the variance of \scarlet's SDR measurements is high and can be larger than the difference between the average for \scarlet and \galsim measurements. After further visual inspection, we realised that in some cases, the SDR for \scarlet reconstructions can be smaller than those of \galsim measurement. An example is shown in Figure \ref{fig:visual_edge} that reveals evident edge truncation, where flux of the galaxy extends to and possibly beyond the edge of the image. It affects both reconstructions, but the \scarlet scheme more so than \galsim. Spatially finite extent is formally the only approximation we make in our scheme because it is practically impossible to extend the summation in equation \ref{eq:whitt2D} over $\mathbb{Z}^2$.
    The problem arises in less than $\sim10\%$ of the images in the HSC to {\it Rubin}, {\it Roman} to HSC and {\it Roman} to {\it Rubin} cases. For the rest of the survey pairs, \galsim is better than scarlet in less than $1\% $ of the cases. We note that the size of the images was fixed at 60 $\times$ 60 pixels for this test, which at least in some cases appears too restrictive to achieve higher SDR, and can be mitigated by additional padding at the expense of increased runtime.

    Figures \ref{fig:SDR} and \ref{fig:SDR_diff} can be reproduced using the notebook \href{https://github.com/herjy/Multi_resolution_comparison/blob/master/Reconstruction_test/Reconstruction_comparison.ipynb}{\faGithub Re\-construction\_comparison}\footnote{https://github.com/herjy/Multi\_resolution\_comparison/ \\ blob/master/Reconstruction\_test/Reconstruction\_comparison.ipynb}. Figures \ref{fig:visual} and \ref{fig:visual_edge} can be reproduced using the \href{https://github.com/herjy/Multi_resolution_comparison/blob/master/Reconstruction_test/Galsim_comparison.ipynb} {\faGithub Galsim\_comparison}\footnote{https://github.com/herjy/Multi\_resolution\_comparison/ \\ blob/master/Reconstruction\_test/Galsim\_comparison.ipynb} notebook.
    
\section{Joint deblending}
\label{Sect:real_images}

    A primary application for our resampling method is the joint modeling and deblending of galaxies by simultaneously fitting the imaging data of multiple surveys. We refer to Melchior et al. (in prep.) for a review of astronomical blending and \citet{Chary2020-rr} for a recent overview of the scientific benefits and operational considerations of joint survey processing.
    
    Specifically, we model multiple galaxies in simulated {\it Euclid} and {\it Rubin} images with \scarlet. In brief, \scarlet assumes that a given patch of the sky contains a predefined number $K$ of sources, each of which is described by a broad-band SED, $A_k$, and a non-parametric model of the morphology, $S_k$. This yields a model
    \begin{equation}
        M = \sum_k^K A_k^\top \times S_k = A\cdot S 
    \end{equation}
    of the data in every band. For multi-instrument modeling, the vector $A_k$ contains the concatenation of the amplitudes of source $k$ in all bands of each available survey.
    It is model $M$, whose single-band images $M_b$ will act as the high-resolution image $I_1$ in equation \ref{eq:A_2}, which is resampled and convolved to the resolution and PSF of the respective surveys. The model parameters $A$ and $S$ are then iteratively optimized by adaptive constrained gradient descent \citep{Melchior2019}. Further details on \scarlet can be found in \citet{Melchior2018}.
    
    \subsection{Simulations setup}

    The simulations were produced with the {\tt galsim} package. Each {\it Rubin} stamp has $60\times60$ pixels, which corresponds to $118\times118$ pixels for the {\it Euclid} images. We inserted between 1 and 10 galaxies in each stamp. The positions of galaxies within the stamp are uniformly distributed, their colours are drawn from the catalog simulation framework \href{https://www.lsst.org/scientists/simulations/catsim}{Catsim}\footnote{https://www.lsst.org/scientists/simulations/catsim}, also used in the construction of the Cosmos DC2 simulations \citep{Korytov2019}. This catalog was generated for the \href{https://github.com/LSSTDESC/BlendingToolKit}{\faGithub BlendingTool\-Kit}\footnote{https://github.com/LSSTDESC/BlendingToolKit} package and can be downloaded at \href{https://stanford.app.box.com/s/s1nzjlinejpqandudjyykjejyxtgylbk}{this address}\footnote{https://stanford.app.box.com/s/s1nzjlinejpqandudjyykjejyxtgylbk}. We artificially draw random magnitudes from a uniform distribution between 20 and 29 for every galaxy. While unrealistic, this step ensures that we can explore the full range of relative blending impacts with a similar number of sources.
    
    Galaxy morphologies are drawn from \galsim's COSMOS catalog of real galaxy images. These galaxy images are convolved with a gaussian kernel with a  $2$-pixel-wide full width at half maximum. This operation ensures that the ground truth galaxies are band-limited. Galaxy positions are drawn at random across the stamps. Individual galaxy images in a given stamp are summed up together before adding noise. We use Gaussian profiles as PSFs with the appropriate width for both surveys and band wavelength range. The details of the parameters used in the simulations are in table \ref{tab:sim_param}. 
    
    \begin{table*}[]
        \centering
        \begin{tabular}{l||cccccc|c}
        &\multicolumn{6}{|c|}{\it Rubin} & {\it Euclid} \\
        bands & u & g & r & i & z & y & VIS \\
        \hline
        Pixel scale (arcsec) & 0.2 & 0.2 & 0.2 & 0.2 & 0.2 & 0.2 & 0.101 \\
        PSF width (arcsec) & 0.327 & 0.31 & 0.297 & 0.285 & 0.276 & 0.267 & 0.16 \\
        exposure time (s) & 1680 & 2400 & 5520 & 5520 & 4800 & 4800 & 2260\\
        zero point & 9.16 & 50.70 & 43.70 & 32.36 & 22.68 & 10.58 & 6.85\\
        sky level & 22.9 & 22.3 & 21.3 & 20.5 & 19.6 & 18.6 & 22.9
        
        \end{tabular}
        \caption{Parameters used to generate simulations of {\it Euclid} and {\it Rubin} blended scenes.}
        \label{tab:sim_param}
    \end{table*}
    
    Each individual galaxy image is saved for comparison with the models reconstructed by \scarlet. For consistency in the comparison, each image was normalised to maximum pixel value of 1 when saved. 
    
    In the resulting simulations, we sacrificed the realism of the scenes to instead homogeneously probe a large number of configurations in terms of magnitudes across bands, galaxy numbers and morphologies.
    
    \subsection{Deblending accuracy}
    
    Deblending is a complex procedure that depends on accurate detection and, in the case of iterative non-linear solvers, initialization of every source.
    In this paper, we are only interested in resampling and the new multi-instrument modelling capabilities it enables. Therefore, we use the ground truth for all source positions and initialise each galaxy morphology model with the  S\'ersic fit to each galaxy image from \galsim. The spectra of each galaxy is initialized through \scarlet, by solving the linear inverse problem of estimating  spectra knowing the initial morphology. We will discuss realistic detection and initialisation schemes for joint-survey analyses in a forthcoming paper. We generated 11625 galaxy profiles in 2315 patches as described above. 
    
   We model each galaxy's morphology with \scarlet's {\tt Ex\-tendedSource} class, i.e. with an image whose pixels satisfy positivity and monotonicity constraints as described in \citep{Melchior2018}. The models maintain a minimal PSF $p_1$ with a Gaussian profile of standard deviation 0.8 pixels in the respective frame.
    
    For the comparison, the true galaxy profiles are rendered with \galsim on the {\it Euclid} and {\it Rubin} grids with their respective pixel integrations and convolved with $p_1$. We then compute the SDR between the \scarlet model and the true image from \galsim. Each stamp is modeled in three different ways: 
    \begin{itemize}
        \item {\bf low-resolution only}: All six {\it Rubin} images are modeled at the {\it Rubin} resolution.
        \item {\bf high-resolution only}: The {\it Euclid} VIS image is modeled at the {\it Euclid} VIS resolution.
        \item {\bf joint modeling}: Both {\it Rubin} and {\it Euclid} images are modelled simultaneously at the {\it Euclid} resolution, using our resampling scheme to match the {\it Rubin} data.
    \end{itemize}
    
    \begin{figure*}[h!]
        \centering
        \includegraphics[ width=\linewidth]{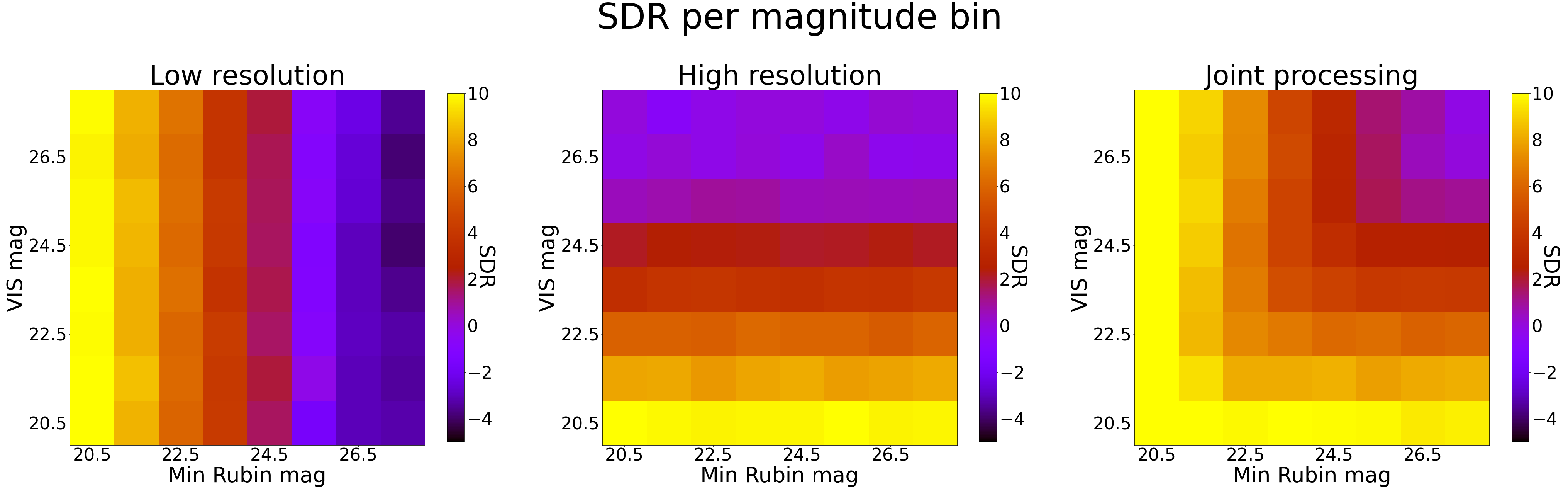}
        \caption{SDR of all modeled galaxies compared to the ground truth as a function of magnitude in {\it Euclid} and {\it Rubin}. Each pixel represents a magnitude bin. The labels on the axis shows the minimum magnitude in the corresponding bin. }
        \label{fig:sdr_per_mag}
    \end{figure*}
    
    In figure \ref{fig:sdr_per_mag}, we show the SDR of the reconstructed images as a function of the galaxies' magnitude in each pair  of {\it Rubin-Euclid} images. Given that {\it Rubin} data have multiple bands, the magnitude used as a reference in these figures is the smallest magnitude for each galaxy across all bands. This lets us probe the maximum impact of using brighter images in the joint resolution modeling. By probing uniformly the range of magnitude from 20 to 30 across all bands and resolutions, we voluntarily create a fraction of images with unrealistic photometry. Given the spectral overlap between {\it Euclid} and {\it Rubin}'s bands, it is unlikely that a galaxy could be very faint in {\it Rubin} images and very bright in {\it Euclid} (bottom right corners of figure \ref{fig:sdr_per_mag}). We choose to leave these unrealistic cases in our study because they let us probe the advantage of combining different surveys with different wavelength ranges, where larger magnitude differences could occur.
    It is evident that each survey individually produces models that are most accurate whenever the sources are bright in that survey's bands. However, joint modeling yields high accuracy when the sources are bright in \emph{any} of the surveys.
    
\begin{figure*}[h!]
        \centering
        \includegraphics[ clip,  width=\linewidth]{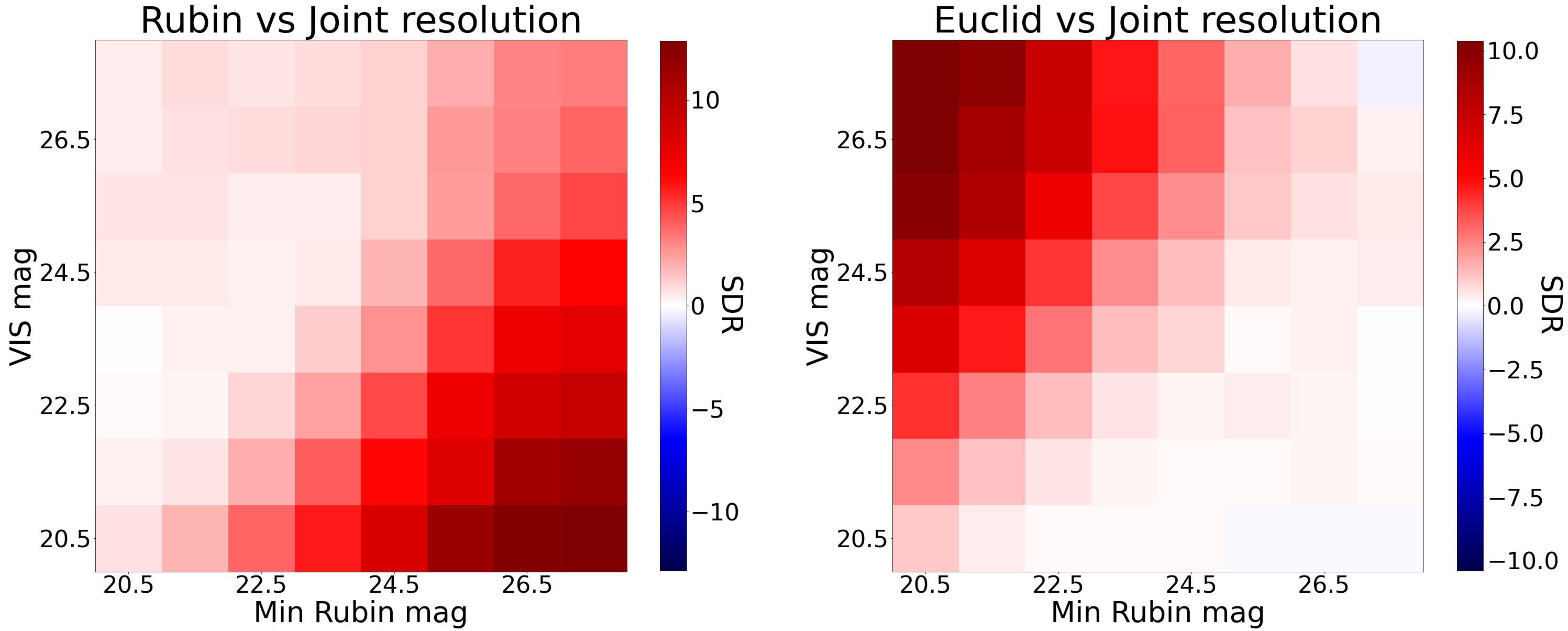}
        \caption{Improvement in SDRs between the single-resolution case (left: {\it Rubin} only; right: {\it Euclid} only) and joint modeling of both observations with \scarlet.}
        \label{fig:single_vs_joint}
    \end{figure*}

    For better interpretation, we present the difference between the SDRs of the single-resolution cases and the joint modeling in figure \ref{fig:single_vs_joint}. The fact that virtually all magnitude bins are colored in red shows that joint modeling yields globally superior reconstruction accuracy. Unsurprisingly, the biggest improvements, of up to an order of magnitude, are from cases where one of the surveys has images that are substantially brighter than the other; e.g. a bright blue source that is easily observed by {\it Rubin}, but not by {\it Euclid}'s redder VIS filter. We want to point out that, while unrealistic in its design, this test provides for the first time clear evidence of the improvement in galaxy modeling from joint pixel-level analysis of blended galaxies.
    
    \hide{
    A closer look at these plots helps understanding exactly the advantage from joint resolution. In figure \ref{fig:sdr_projection}, we show the results presented in figure \ref{fig:sdr_per_mag} average along the different resolutions. The left-most panel of figure \ref{fig:sdr_projection} shows the the average SDR per mean magnitude bin. We see that joint resolution provides better SDR in the lower magnitudes. At low magnitudes, the SDR of joint resolution matches the average SDR from {\it Euclid}'s resolution, meaning that joint resolution takes advantage of the best of both resolutions. The same effect can be observed in the right-most panel where, at high VIS magnitude, joint resolution SDR matches the SDR from the averaged low resolution SDRs. 
    
    \begin{figure*}[h!]
        \centering
        \includegraphics[trim = 0cm 0cm 0cm 0cm, clip,  scale=0.2]{Figures/SDR(Rubin_mag).png}
        \includegraphics[trim = 0cm 0cm 0cm 0cm, clip,  scale=0.2]{Figures/SDR(Euclid_mag).png}
        \caption{{\it left panel}: shows the panels from figure \ref{fig:sdr_per_mag} from left to right averaged along the VIS mag axis in green, blue and red respectively.{\it right panel}: shows the panels from figure \ref{fig:sdr_per_mag} from left to right averaged along the {\it Rubin} mag axis in green, blue and red respectively. }
        \label{fig:sdr_projection}
    \end{figure*}
    }
    
    All the figures presented in this section can be reproduced from the notebooks found at  \href{https://github.com/herjy/Multi_resolution_comparison/tree/master/Reconstruction_test}{\faGithub multi-resolution simulations}\footnote{https://github.com/herjy/Multi\_resolution\_comparison/ \\ tree/master/Reconstruction\_test}.
    
\section{Conclusion}
\label{Sect:conclusion}
    We introduce a novel method to express convolution and resampling of one- and two-dimensional functions as a single operation directly based on the Whittaker-Shannon interpolation formula. Applied to astronomical images it allows one to render a high-resolution image with a given PSF to a lower-resolution image with a different PSF. The assumptions for this method are that the high-resolution image is critically sampled, the amount of light at the edge of the high-resolution image is negligible, and the PSF of the high-resolution image is narrower than that of the low-resolution image.
    
    We compare our resampling scheme with the method by \citet{Bernstein2014}, implemented in \galsim, which uses a much more compact approximated interpolation kernel. 
    We find that, with the accelerated version of our scheme, the runtimes of the resampling and convolution operation are comparable for very small images of $50\times 50$ pixels and, for image of $200\times200$ pixels, a factor of $\sim 10$ slower. The most costly operation is a matrix multiplication, for which GPU architectures should provide substantial performance gains. However, we consistently find more accurate interpolation results, with residuals typically being reduced by an order of magnitude compared to \galsim.
    
    We implement this method in the galaxy modeling framework \scarlet where it enables us to simultaneously fit images from different instruments. To our knowledge, this work presents the first application of non-parametric multi-band modeling across multiple instruments.
    We apply this new capability to the pressing problem of deblending deep astronomical imaging surveys, specifically those planned with the {\it Euclid} mission and the {\it Vera C. Rubin Observatory}.
    We find that joint modeling increases the reconstruction accuracy, as quantified by the source distortion ratio SDR of the morphological model, by up to an order of magnitude compared to analyzing both survey images separately. This study is, to our knowledge, the first investigation of the gains of joint pixel-level analysis for galaxies with a wide range of blending configurations. Follow-up studies can use our codes to investigate 
    the accuracy of joint reconstructions for other quantities of interest, such as shape, photometry, and colour measurements.
    Simulation efforts to produce highly realistic images for upcoming surveys \citep[e.g.]{Troxel2019, Korytov2019} will be instrumental for reaching this goal.
    
    \hide{
    One of the added values of the resampling method presented here is the power to perform efficient resampling in higher dimensions. Indeed, the technique of splitting sinc interpolation along different directions between the two argument of the convolution can be further employed in higher dimensions by applying all the directional sinc convolutions but one  to the difference kernel, thus bounding the complexity of the gradient step operation.
    } 
    
    In forthcoming works, we will continue the outlined path towards practically feasible pixel-level analysis of multiple surveys by addressing aspects we ignored in this work, in particular the problem of source detection and association across multiple instruments, the impact of astrometric and photometric calibration errors, and the initialisation and modeling of complex morphological patterns.

\section{Acknowledgements}

    This paper makes use of software developed for the Large Synoptic Survey Telescope. We thank the LSST Project for making their code available as free software at  \url{http://dm.lsst.org}.
    
    This material is based upon work supported by the National Science Foundation under  Cooperative  Agreement  1258333  managed  by  the  Association  of  Universities for Research in Astronomy (AURA), and the Department of Energy under Contract  No.  DE-AC02-76SF00515 with  the  SLAC  National  Accelerator  Laboratory.  Additional funding  for  {\it Rubin}  Observatory  comes  from  private  donations, grants to universities, and in-kind support from LSSTC Institutional Members.
    
    The \scarlet framework is written in {\tt python} with {\tt C++} extensions accessible through {\tt pybind11} \citep{pybind11} and makes use of the {\tt numpy} \citep{VanDerWalt2011}, {\tt scipy} \citep{Virtanen2020} and {\tt astropy} \citep{astropy2018}, \href{https://github.com/HIPS/autograd}{autograd} and \href{https://github.com/pmelchior/proxmin}{proxmin}  packages. The figures presented in this paper were produced using the {\tt matplotlib} package \citep{Hunter2007}.
    
    The results presented in this paper are reproducible using three packages developed by the authors. The versions of these packages as used in this publication are the following: \href{https://github.com/pmelchior/scarlet}{\faGithub \scarlet}\footnote{https://github.com/pmelchior/scarlet} commit (f78a044), \href{https://github.com/herjy/scarlet_extensions}{\faGithub scarlet\_extensions}\footnote{https://github.com/herjy/scarlet\_extensions/releases/tag/paper}  and \href{https://github.com/herjy/Multi_resolution_comparison}{\faGithub Multi\_resolution\_comparison}\footnote{https://github.com/herjy/Multi\_resolution\_comparison/releases/tag/paper}.
    
\section*{References}

\bibliography{mybibfile}

\newpage
\appendix

\end{document}